\documentclass[12pt]{article}
\usepackage{amsmath}
\usepackage{amsfonts}
\usepackage{amssymb}
\usepackage{ifthen}
\usepackage{color}

\allowdisplaybreaks[4]     

\newcommand{\showlabel}[1]{
  \label{#1}
}

\newcommand{\lra}{\leftrightarrow}

\newcommand{\nn}{\nonumber}
\newcommand{\er}[1]{(\ref{#1})}          
\renewcommand{\L}{Lema\^{\i}tre}
\newcommand{\LT}{{\L}-Tolman}
 
\newcommand{\mb}{\mathbf}
\newcommand{\bs}[1]{\mbox{\boldmath $#1$}}

\newcommand{\be}{{\bf e}}     
\newcommand{\bp}{{\bs{\p}}}     
\newcommand{\bd}{{\bs{\d}}}     
\newcommand{\bv}{{\bf v}}     

\newcommand{\Rt}{\dot{R}}

\newcommand{\sth}{\,\sin\theta\,}
\newcommand{\cth}{\,\cos\theta\,}
\newcommand{\sph}{\,\sin\phi\,}
\newcommand{\cph}{\,\cos\phi\,}

\renewcommand{\d}{{\rm d}}     
\newcommand{\p}{\partial}     
\newcommand{\td}[2]{\frac{{\rm d} #1}{{\rm d} #2}}     
\newcommand{\pdil}[2]{\partial #1/\partial #2 }     

\begin{document}
\sffamily

\title{Frame Rotation in the Szekeres Spacetimes}
 
\author{
  Charles Hellaby
  \thanks{\tt Charles.Hellaby@uct.ac.za} \\
  {\small \it Dept. of Maths. and Applied Maths,
  University of Cape Town,
  Rondebosch,
  7701,
  South Africa}
}
 
\date{}

 
\maketitle

\begin{abstract}
The Szekeres metric is an inhomogeneous cosmological model without any symmetries.  The standard Riemann-type coordinates can be transformed into spherical-type coordinates, but the metric is no longer diagonal, and the constant ``radius" 2-spheres, 2-hyperboloids or 2-planes are known to be ``non-concentric".  Since the transformation into spherical-type coordinates is ``radius" dependent, we question whether these coordinates have the same orientation on each 2-surface.  To answer this question, we set up an orthonormal tetrad (ONT), and investigate its variation.  We find that a relative rotation of the tetrad is generic, and it can increase systematically under conditions that are not very restrictive.  We search for paths along which the tetrad is constant, and find they only exist under very restrictive conditions.  In the process, we create a systematic method for defining an ONT with chosen properties from a given metric.
\end{abstract}

 
\section{Motivation}

The Szekeres (S) metric is an exact inhomogeneous solution of the EFEs that has no Killing vectors.  It has 6 free functions of ``radial" coordinate $r$, plus the parameter $\epsilon$.  However the 2-surfaces of constant $(t, r)$ are spheres, planes, or pseudo-spheres (2-sheeted right hyperboloids of revolution), depending on whether the parameter $\epsilon$ is $+1$, $0$, or $-1$.  It is well known that the constant $(t, r)$ shells are ``non-concentric", i.e. the separation between adjacent constant $r$ shells depends on the remaining coordinates $p$ \& $q$, which are related to standard ``angular" coordinates $\theta$ \& $\phi$ by a Riemann projection.%
\footnote{\sf Terms like ``radial", ``angle" and ``non-concentric" have pseudo-spherical equivalents \cite{HelKra08}.}
 However, given that the $(p,q)$ to $(\theta,\phi)$ transformation is dependent on $r$, one may wonder whether ``non-concentricity" is the whole story; do adjacent shells have constant ``orientation" of the $\theta$-$\phi$ frame?  It would appear that this has been tacitly assumed up to now, but it would be useful for our understanding to have this assumption confirmed or disproved.  In this paper we focus on the $\epsilon = +1$ case, adding the key formulas for the other cases in appendices.

More precisely, our main question is, given a constant $t$ slice and a particular $(\theta,\phi)$ angle, do the tangents \& normals to successive 2-surfaces of constant $r$, at this angle, all point in the same direction?  Of course, there is no such thing as a globally constant vector in a curved space(time).  Nevertheless, the Friedmann-{\L}-Robertson-Walker (FLRW) \& {\LT} (LT) spacetimes do have angular directions with constant 3-d ``orientations"; along constant-time radial paths, the radial and angular basis vectors do remain 3-parallel (see example (d) of appendix \ref{ONTVarExmpls}).  If, compared with such spherically symmetric behaviour, the S metric shows large or systematic changes, this would be significant.  

This primary question leads to several others.  If we find that the radial variation of orientation is not zero, then we would next ask whether the evolution in time also changes angular orientations in 3-d.  In addition, in a given S metric (with specified arbitrary functions) are there paths along which there is no variation in the orientation of $\theta$ and $\phi$?  If there are, this would be an inhomogeneous version of the constant orientation along radial paths of spherically symmetric metrics.  If not, what are the conditions on the S arbitrary functions that would ensure no variation in orientation?

Plots of sections through S models are problematic because of curvature; one has to choose a slicing, and then a projection.  Even with a 3-space of constant curvature, representations of say the density distribution on a 2-surface are necessarily distorted, due to the impossibility of mapping a curved 2-space onto flat paper without some distortion of distances and angles.  Thus a discovery that the $(\theta,\phi)$ coordinates do not preserve orientation would add to the complexity of representing the S metric graphically.

\section{The Szekeres Spacetime}

The Szekeres (S) metric \cite{Szek75a, Szek75b} is a non-symmetric generalisation of the {\LT} (LT) metric \cite{Lem33, Tol34} and the Ellis metrics \cite{Elli67}, in which the shells of constant $t$ and $r$ are arranged ``non-concentrically".  The line element is
\begin{align}
   ds^2 = - \d t^2 + \frac{\left( R' + \frac{R E'}{E} \right)^2 \, \d r^2}{W^2} + \frac{R^2}{E^2} \left( dp^2 + dq^2 \right) ~,
   \showlabel{ds2Sz}
\end{align}
where $W = \sqrt{\epsilon + f}\;$, ${}'$ indicates $\pdil{}{r}$, $R = R(t,r)$ is a kind of areal factor that gives the ``size" of the constant $t$ \& $r$ 2-surfaces, and $f = f(r)$ is an arbitrary function that determines the {\it local\/} 3-geometry: positively and negatively curved for $f < 0$ and $f > 0$, and flat for $f = 0$.  The factor
\begin{align}
   E & = \frac{S}{2} \left( \frac{(p - P)^2}{S^2} + \frac{(q - Q)^2}{S^2} + \epsilon \right) ~,
\end{align}
contains the 3 arbitrary functions $S = S(r)$, $P = P(r)$ and $Q = Q(r)$, and $\epsilon$ has the 3 values $+1, 0, -1$ which determine whether each of the constant $r$ surfaces have a spherical, planar, or right-hyperboloidal (pseudo-spherical) 2-geometry respectively.  Although the individual 2-surfaces of constant $t$ \& $r$ have constant curvature, they are not arranged symmetrically, so the 3-spaces of constant $t$ and the 4-metric do not have spherical, planar, or hyperboloidal symmetry.  It is entirely possible for a single model to have $f$ ranging from $-$ve to $+$ve and/or all 3 signs for $\epsilon$.  Note however that not all combinations of signs of $f$ \& $\epsilon$ are possible \cite{HelKra08}.  The model evolution is determined by 
\begin{align}
   \Rt^2 = \frac{2 M}{R} + f + \frac{\Lambda R^2}{3} ~,   \showlabel{RtSq}
\end{align}
which is a generalised Friedmann equation, and here $\Lambda$ is the cosmological constant, while $\dot{}$ indicates $\pdil{}{t}$.  When solving the $\Lambda = 0$ version of \er{RtSq}, we see that if $f < 0$ the evolution is re-collapsing or ``elliptic", if $f > 0$ it is montonic expansion (or collapse) i.e. ``hyperbolic", and if $f = 0$ we have the borderline ``parabolic" case.  Thus $f$ has a second interpretation as an energy parameter, while the arbitrary function $M = M(r)$ represents a mass-like parameter in a term $2M/R$ that looks like a gravitational potential energy.  In quasi-spherical regions it is indeed the gravitational mass interior to the sphere of constant $r$.  The last arbitrary function, $a = a(r)$, appears in the integration of \er{RtSq}, and gives the time of the big bang, $t = a(r)$, when $R(a(r), r) = 0$, on each $r$ shell's ``world sheet".  The matter is pressure-free and comoving,
\begin{align}
   u^a & = \delta^a_t ~,   \showlabel{comov}
\end{align}
and the density is 
\begin{align}
   \kappa \rho & = \frac{2 (M' - 3 M E' / E)}{R^2 (R' - R E' / E)} ~.   \showlabel{rho}
\end{align}

For further information, see the excellent survey of inhomogeneous cosmologies \cite{Kra97}, and also \cite{Hel09}, \cite{PleKra06} and \cite{BoKrHeCe10}.  
Some important geometric and physical results are that the metric has no killing vectors \cite{BoSuTo77}, the constant $t$ sections are conformally flat \cite{BeEaOl77}, there is no gravitational radiation \cite{Bonn76b,Cova80}, the null limit of the metric is a generalisation of the Kinnersley rocket \cite{Hell96b}, and the Datt-Kantowski-Sachs type S model ($\beta' = 0$) is a limit of the more common \LT\ type ($\beta' \neq 0$) \cite{Hell96b}.  
Much interesting work has been done using this metric, \cite{
Bonn76b, 
BonTom76, 
BeEaOl77, 
BoSuTo77, 
Wain77, 
Cova80, 
GooWai82a, 
GooWai82b, 
BarSte84, 
Glei84, 
deS85, 
Bonn86, 
BonPug87, 
Hell96b, 
JosKro96, 
HelKra02, 
Bole06a, 
ApoCar07, 
Bole07, 
NolDeb07, 
ChaDeb08, 
HelKra08, 
IRGWNS08, 
Kra08, 
Bole09a, 
Bole09b, 
BolWyi09, 
BolCel10, 
Bole11, 
BolSus11, 
Debn11, 
KraBol11, 
MeuBru11, 
NwIsTh11, 
HePrIbCa12, 
IshPee12, 
KraBol12a, 
KraBol12b, 
MiCeSi12, 
PeIsTr12, 
SusBol12, 
WalHel12, 
MisCel14, 
PeTrIs14, 
Vill14, 
VrbSvi14, 
KokHan15a, 
KokHan15b, 
SusGas15, 
Apos16b,
BolNazWil16, 
Kra16, 
SuGaHi16, 
Apos17a, 
GeoHel17,
Koks17,
SuHiGaGe17%
}.

\subsection{Polar Coordinates for the S Metric}


The transformations from $(p, q)$ coordinates to polar-type coordinates are the stereographic projections:
\begin{align}
   \epsilon & = +1~:~~~~~~~~ &
   p & = P + S \cot\left(\frac{\theta}{2}\right) \cph ~, &~~~~
   q & = Q + S \cot\left(\frac{\theta}{2}\right) \sph ~,
   \showlabel{RmPrS} \\[2mm]
   \epsilon & = ~0~: & 
   p & = P + S \left(\frac{2}{\theta}\right) \cph ~, &~~~~
   q & = Q + S \left(\frac{2}{\theta}\right) \sph ~,
   \showlabel{RmPrP} \\[2mm]
   \epsilon & = -1~: &
   p & = P + S \coth\left(\frac{\theta}{2}\right) \cph ~, &~~~~
   q & = Q + S \coth\left(\frac{\theta}{2}\right) \sph ~.
   \showlabel{RmPrH}
\end{align}
For the quasi-spherical regions of the S metric \er{ds2Sz}, where $\epsilon = +1$, the term $(\d p^2 + \d q^2)/E^2$ is actually a unit 2-sphere, and the $(p, q)$ coordinates are understood to be Riemann projections of normal angular coordinates.  Regions where $\epsilon = -1$ are Lobachevsky-Bolyai 2-surfaces of constant negative curvature, and the above transformation maps a Minkowski hyperboloid or Lorentz hyperboloid onto the Poincare disc%
\footnote{\sf The region outside this disc maps to the second sheet of the hyperboloid, but \cite{HelKra08} shows that only one of the two hyperboloid sheets can be made regular.}%
, the region within $(p - P)^2 + (q - Q)^2 = S^2$.  Where $\epsilon = 0$, the $(p, q)$ 2-surfaces are planes, and the above transformation is an inversion in the same circle.

We here present the equations for the quasi-spherical S model, $\epsilon = + 1$, while the main equations for quasi-hyperboloidal and quasi-planar models are collected in appendices \ref{AppEp-1} and \ref{AppEp0}.  They are qualitatively similar.

By \er{RmPrS}, the $\epsilon = +1$ metric in $(\theta, \phi)$ coordinates is
\begin{align}
   g_{tt} & = - 1 ~,~~~~
   g_{\theta\theta} = R^2 ~,~~~~
   g_{\phi\phi} = R^2 \sin^2\theta ~, \nn \\
   g_{rr} & = \frac{\big(R' S + R [S' \cos\theta + \sin\theta \{P' \cos\phi + Q' \sin\phi\}]\big)^2}{S^2 W^2} \nn \\
      &~~~~ + \frac{R^2}{S^2} \big(S' \sin\theta + (1 - \cos\theta) \{P' \cos\phi + Q' \sin\phi\}\big)^2 \nn \\
      &~~~~ + \frac{R^2}{S^2} (1 - \cos\theta)^2 \{P' \sin\phi - Q' \cos\phi\}^2 ~, \nn \\
   g_{r\theta} & = - \frac{R^2 (S' \sin\theta + (1 - \cos\theta) \{P' \cos\phi + Q' \sin\phi\})}{S} \nn \\
   g_{r\phi} & = - \frac{R^2 (1 - \cos\theta) \sin\theta \{P' \sin\phi - Q' \cos\phi\}}{S} ~.
   \showlabel{SzThPhMetric}
\end{align}
Thus the coordinates are not orthogonal, and the off-diagonal components depend on both $r$ \& $t$.  The functions $E$ and $E'$ become
\begin{align}
   E = \frac{S}{(1 - \cth)} ~,~~~~~~ 
   E' = \frac{\cth S' + \sth (P' \cph + Q' \sph)}{(1 - \cth)} ~.
\end{align}

\subsection{The Szekeres ONT in $t,r,\theta,\phi$ Coordinates}
\showlabel{SzAngONT}

Here we use the formulae of appendix \ref{FndONT} to determine an ortho-normal tetrad (ONT) for the angular Szekeres metric \er{SzThPhMetric}.  We will write the ONT labels, and any indices in that basis, with brackets round them.  We specify that the $\be_{(t)}$, $\be_{(\theta)}$ \& $\be_{(\phi)}$ basis vectors line up with the $t$, $\theta$ \& $\phi$ coordinate directions, and then define $\be_{(n)}$ as the 4th one, automatically orthogonal to the other three.%
\footnote{\sf We don't write $\be_{(r)}$.  In $(r, \theta, \phi)$ coordinates, the $r$ coordinate is not orthogonal to $\theta$ \& $\phi$ directions.  In $(r, p, q)$ coordinates, though, the spatial coordinates are all orthogonal.}  
Potentially these directional constraints give 9 equations, which is 3 too many, but if we specify the 6 constraints
\begin{align}
   \be_{(t)}(\bd r) = 0 = \be_{(t)}(\bd \theta) = \be_{(t)}(\bd \phi) ~,~~~~
   \be_{(\theta)}(\bd r) = 0 = \be_{(\theta)}(\bd \phi) ~,~~~~
   \be_{(\phi)}(\bd r) = 0 ~,   \showlabel{SzONTdirnconstr}
\end{align}
then $\be_{(\theta)}(\bd t) = 0$, $\be_{(\phi)}(\bd t) = 0$, $\be_{(\phi)}(\bd \theta) = 0$ follow automatically from the fact that $g_{t\theta} = 0 = g_{t\phi} = g_{\theta\phi}$.

Solving \er{ONT} \& \er{SzONTdirnconstr}, using GRTensor \cite{GRT} \& a Maple \cite{Maple} worksheet, the ONT for the $\epsilon = +1$ angular Szekeres metric \er{SzThPhMetric} has the following non-zero components
\begin{align}
   e_{(a)}{}_i & = 
   \begin{pmatrix}
   -1 & 0 & 0 & 0 \\
   0 & \dfrac{1}{W} \left( R' + \dfrac{R [S' \cos\theta + \sin\theta \{P' \cos\phi + Q' \sin\phi\}]}{S} \right) & 0 & 0 \\[4mm]
   0 & - \dfrac{R [S' \sin\theta + (1 - \cos\theta) \{P' \cos\phi + Q' \sin\phi\}]}{S} & R & 0 \\[3mm]
   0 & - \dfrac{R (1 - \cos\theta) \{P' \sin\phi - Q' \cos\phi\}}{S} & 0 & R \sin\theta
   \end{pmatrix}
   \showlabel{SzThPhONT}
\end{align}
where rows 1 to 4 give the components of $\be_{(t)}$, $\be_{(n)}$, $\be_{(\theta)}$, $\be_{(\phi)}$, respectively.  It may be verified that the above basis satisfies all 19 of our requirements.  As it turns out, the $e_{(n)}{}_r$ component has simplified greatly because of extensive cancellation in $g_{rr} - (g_{r\phi}^2 g_{\theta\theta}^{} + g_{r\theta}^2 g_{\phi\phi}^{})/(g_{\theta\theta}^{} g_{\phi\phi}^{})$.  Put another way, the extra terms in $g_{rr}$ are generated by $e_{(\theta)}{}_r$ and $e_{(\phi)}{}_r$.

\section{Variation of the Basis Vectors}

For this investigation, we must look at how the basis vectors in the appropriate directions vary.  However, the $(r, \theta, \phi)$ coordinate basis vectors are non-orthogonal as well as having non unit magnitude.  To obtain more transparent results, it is sensible to use the orthonormal tetrad (ONT) derived above.  The connection that defines their variation is, for an ONT, the set of Ricci rotation coefficients,
\begin{align}
   \nabla_{\be_{(a)}} \be_{(b)} & = \Gamma^{(c)}{}_{{(a)}{(b)}} \be_{(c)} ~~~~~~\lra~~~~~~
   \big( \nabla_{\be_{(a)}} \be_{(b)} \big) \be^{(c)} = e_{(a)}{}^d \big( \nabla_d e_{(b)}{}^f \big) e_f{}^{(c)}
      = \Gamma^{(c)}{}_{{(a)}{(b)}} ~.
      \showlabel{RRCeq}
\end{align}
The variation we are interested in, is the one along a particular path with tangent vector $\bv = v^j \be_j$,
\begin{align}
   {\cal V}^{(c)}{}_{(b)} \equiv \big( \nabla_{\bv} \be_{(b)} \big) \be^{(c)} & = v^{(a)} \, \Gamma^{(c)}{}_{{(a)}{(b)}}
      = v^j e^{(a)}{}_j \, \Gamma^{(c)}{}_{{(a)}{(b)}} ~,
      \showlabel{RRCv}
\end{align}
where ${\cal V}^{(c)}{}_{(b)}$ is being defined here.%
\footnote{\sf Our notation for commutation coefficients and connection components is
\begin{align*}
   \gamma^{(c)}{}_{(a)(b)} & = [ \be_{(a)}, \be_{(b)}](\be^{(c)})
      = \{ (e_{(a)})^i (e_{(b)})^j{}_{,i} - (e_{(b)})^i (e_{(a)})^j{}_{,i} \} (e^{(c)})_j
      = \{ (e^{(c)})_{i,j} - (e^{(c)})_{j,i} \} (e_{(a)})^i (e_{(b)})^j ~, \\
   \Gamma_{(c)(d)(e)} & = \frac{1}{2} \{ g_{(c)(d),(e)} + g_{(e)(c),(d)} - g_{(d)(e),(c)} \}
      + \frac{1}{2} \{ \gamma_{(c)(d)(e)} + \gamma_{(e)(c)(d)} - \gamma_{(d)(e)(c)} \} ~.
\end{align*}
}
We don't want to use ONT components of $v^j$, because we want the rate of change of the ONT with respect to a parameter along a chosen path.  Particular paths will be specified later.

\subsection{Obtaining the Rates of ``Rotation" \& ``Boost" of the Basis}

Now for an ON basis, the rate of variation of the basis is due to the rate of change of a unitary rotation/boost matrix.  If $\Lambda^{(c)}{}_{(a)}(\lambda)$ is a Lorentz transformation matrix whose components depend on some parameter $\lambda$ along a given path $x^b = x^b(\lambda)$, then its action on some fixed vector $\xi^{(a)}|_0$ (at a given $\lambda_0$) generates a transformed vector that depends on $\lambda$, $\xi^{(c)}(\lambda) = \Lambda^{(c)}{}_{(a)} \, \xi^{(a)}|_0$.  The rate of change of the transformed $\xi^{(b)}$ along the path is then
\begin{align}
   \td{\xi^{(c)}}{\lambda} = \td{\Lambda^{(c)}{}_{(d)}}{\lambda} \, \xi^{(d)}|_0 ~~~~\to~~~~
   \left. \td{\xi^{(c)}}{\lambda} \right|_0 = \left. \td{\Lambda^{(c)}{}_{(d)}}{\lambda} \right|_0 \; \xi^{(d)}|_0
\end{align}
which holds for every $\lambda_0$ value, so we have at each point on the curve
\begin{align}
   \td{\xi^{(c)}}{\lambda} = \td{\Lambda^{(c)}{}_{(d)}}{\lambda} \, \xi^{(d)} ~.
\end{align}
If instead we knew $\Lambda^{(c)}{}_{(d)}$ as a function of position, and we knew the path tangent vector $v^{(d)}$, then we could write
\begin{align}
   \td{\xi^{(c)}}{\lambda} = v^{(a)} \, \p_{(a)} \Lambda^{(c)}{}_{(d)} \, \xi^{(d)} ~.
\end{align}
Further, if we choose $\xi^{(d)}$ to be a unit basis vector $e_{(b)}{}^{(d)} = \delta^{(d)}_{(b)}$, then
\begin{align}
   \td{e_{(b)}{}^{(c)}}{\lambda} = v^{(a)} \, \p_{(a)} \Lambda^{(c)}{}_{(b)}
      = \td{\Lambda^{(c)}{}_{(b)}}{\lambda} ~.
\end{align}
Comparing this with \er{RRCv}, which gives the rate of change of the unit basis vector $\be_{(b)}$, we see that ${\cal V}^{(c)}{}_{(b)} = v^{(a)} \, \Gamma^{(c)}{}_{{(a)}{(b)}}$ is a rate of rotation/boost matrix.

Appendix \ref{RBA} describes the procedure for extracting the rotation axis, the rate of rotation, the boost direction, and the rate of boost, given such a rate of rotation/boost matrix.  It also gives some examples to provide background to the results below.

\subsection{The Questions}

Now the main question is~ (i) whether the $\be_{(n)}$, $\be_{(\theta)}$, $\be_{(\phi)}$ vectors maintain constant orientation along constant $(\theta, \phi)$ paths, relative to the constant $t$ 3-spaces?
(We do of course expect 4-d variation due to expansion, and it is known that even in FLRW spacetimes the constant $t$ 3-spaces are not geodesic.)  For the $r$ coordinate paths, which have constant $(\theta, \phi)$, we merely choose, $v^i = \delta^i_r$ with $\be_{(b)}$ \& $\be^{(c)}$ spatial, and check whether
\begin{align}
   e^{(a)}{}_r \, \Gamma^{(c)}{}_{{(a)}{(b)}} = 0
\end{align}
holds.

Our auxilliary questions are as follows.  (ii) Given a S metric, are there paths along which there is no variation in the orientation of $\theta$ and $\phi$?  In other words, try to solve for $v^i$ such that
\begin{align}
   {\cal V}^{(c)}{}_{(b)} = v^j e^{(a)}{}_j \, \Gamma^{(c)}{}_{{(a)}{(b)}} = 0 ~.   \showlabel{rVbV0path}
\end{align}
(iii) If this is not possibe in general, what are the conditions on the arbitrary functions $S$, $P$, $Q$ that would ensure \er{rVbV0path} holds?
(iv) Does the time evolution cause changes in the 3-space orientation
\begin{align}
    {\cal V}^{(c)}{}_{(b)}\Big|_{\bv = \be_{(t)}} = \big( \nabla_{\be_{(t)}} \be_{(b)} \big) \be^{(c)}
    & = \Gamma^{(c)}{}_{{(t)}{(b)}} \neq 0 ~,
\end{align}
with $(b)$ \& $(c)$ spatial components?  Calculating \er{rVbV0path} allows us to address all these questions.

\section{Variation of the Szekeres Angular ONT}

The ONT for the $\epsilon = +1$ angular S metric is given in \er{SzThPhONT}.  We calculate ${\cal V}^{(c)}{}_{(b)}$ directly from this ONT, working in $t, r, \theta, \phi$ coordinates, using a Maple worksheet and GRTensor%
\footnote{\sf The translation from the present notation to GRTensor notation is
\begin{align*}
   \gamma_{(a)[(b)(c)]} = {}_{_{GRT}}\lambda_{[(b)|(a)|(c)]} = {\tt lambda(bdn,bdn,bdn)} \\
   \Gamma_{[(a)|(b)|(c)]} = {}_{_{GRT}}\gamma_{[(a)(c)](b)} = {\tt rot(bdn,bdn,bdn)}
\end{align*}
}%
.  We give it in terms of the coordinate components $v^j$.
\begin{align}
   {\cal V}^{(n)}{}_{(t)} & = \frac{(\Rt' S + \Rt [S' \cos\theta + \sin\theta \{P' \cos\phi + Q' \sin\phi\}])}
       {S W} \; v^r   \showlabel{Vnt-thph} \\
   {\cal V}^{(\theta)}{}_{(t)} & = \frac{- \Rt [S' \sin\theta + (1 - \cos\theta)\{P' \cos\phi + Q' \sin\phi\}]}{S} \; v^r
      + \Rt \; v^\theta   \showlabel{Vtht-thph} \\
   {\cal V}^{(\phi)}{}_{(t)} & = \frac{- \Rt (1 - \cos\theta)\{P' \sin\phi - Q' \cos\phi\}}{S} \; v^r
      + \Rt \sin\theta \; v^\phi   \showlabel{Vpht-thph} \\
   {\cal V}^{(\theta)}{}_{(n)} & = - \bigg( \frac{S' (W^2 - 1) \sin\theta + [W^2 - \cos\theta (W^2 - 1)]
      \{P' \cos\phi + Q' \sin\phi\}}{W S} \bigg) v^r + W v^\theta   \showlabel{Vthn-thph} \\
   {\cal V}^{(\phi)}{}_{(n)} & = \frac{[1 - W^2 (1 - \cos\theta)]\{P' \sin\phi - Q' \cos\phi\}}{S W} \; v^r
      + W \sin\theta \; v^\phi   \showlabel{Vphn-thph} \\
   {\cal V}^{(\phi)}{}_{(\theta)} & = \frac{- \sin\theta \{P' \sin\phi - Q' \cos\phi\}}{S} \; v^r
      + \cos\theta v^\phi   \showlabel{Vphth-thph}
\end{align}
We have checked that this matrix has the required properties: the symmetric part has zero determinant and is non-zero only for time-space components; the antisymmetric part has zero determinant and is non-zero only for space-space components.  The equivalent equations for the other $\epsilon$ values are in \er{Vnt-thphH}-\er{Vphth-thphH} and \er{Vnt-thphP}-\er{Vphth-thphP}.

The rotation rate and axis are
\begin{align}
   \dot{\alpha} = \sqrt{\big[{\cal V}^{(\phi)}{}_{(\theta)}\big]^2 + \big[{\cal V}^{(\phi)}{}_{(n)}\big]^2
               + \big[{\cal V}^{(\theta)}{}_{(n)}\big]^2}\; ~,~~~~
   a^c = \frac{\left( 0, {\cal V}^{(\phi)}{}_{(\theta)}, -{\cal V}^{(\phi)}{}_{(n)}, {\cal V}^{(\theta)}{}_{(n)} \right)}
              {\dot{\alpha}}
\end{align}
and the boost  rate and axis are
\begin{align}
   \dot{\chi} = \sqrt{\big[{\cal V}^{(n)}{}_{(t)}\big]^2 + \big[{\cal V}^{(\theta)}{}_{(t)}\big]^2
               + \big[{\cal V}^{(\phi)}{}_{(t)}\big]^2}\; ~,~~~~
   b^c = \frac{\left( 0, {\cal V}^{(n)}{}_{(t)}, {\cal V}^{(\theta)}{}_{(t)}, {\cal V}^{(\phi)}{}_{(t)} \right)}
              {\dot{\chi}} ~.
\end{align}

As a further check, if we put $v^r = 0$, we should get the rotations and boosts appropriate to an expanding spherical surface;
\begin{align}\begin{aligned}
   & {\cal V}^{(n)}{}_{(t)} = 0 ~,~~~~~~
   && {\cal V}^{(\theta)}{}_{(t)} = \Rt \; v^\theta ~,~~~~~~
   && {\cal V}^{(\theta)}{}_{(n)} = W \; v^\theta \\
   & {\cal V}^{(\phi)}{}_{(t)} = \Rt \sin\theta \; v^\phi ~,~~~~~~
   && {\cal V}^{(\phi)}{}_{(n)} = W \sin\theta \; v^\phi ~,~~~~~~
   && {\cal V}^{(\phi)}{}_{(\theta)} = \cos\theta v^\phi ~.
\end{aligned}\end{align}
These are identical to those for an LT model given in \er{eVLT}, with rotation and boost rates and axes as given in \er{rotrateaxLT} and \er{bstrateaxLT}.  This means that ONT rotations within each evolving 2-sphere are fully understood.  See appendix \ref{RBA} for the interpretation.

\subsection{Analysis of the Cases}
 
We see immediately that the rotation rate has no dependence on $v^t$, so as expected the ONT does not rotate along the comoving matter worldlines.

Next consider a pure $r$ direction with $r$ the path parameter, i.e. $0 = v^t = v^\theta = v^\phi$, and $v^r = 1$.  By definition, $\theta$ \& $\phi$ (\& $t$) are constant along $r$ coordinate lines, so one might expect the tetrad to have constant orientation, as it does in the LT model, but this is not so.  

For the sub-case of $0 = P' = Q'$, the relative rotation is in the $(r, \theta)$ plane, 
\begin{align}\begin{aligned}
   \dot{\alpha} & = \frac{S' (1 - W^2) \sin\theta}{S W} v^r ~,~~~~~~
   a^i = \delta^i_\phi ~.   \showlabel{Pr0Qr0}
\end{aligned}\end{align}
(See \er{alpha-dot-Pr0Qr0H} \& \er{alpha-dot-Pr0Qr0P} for $\epsilon = -1,0$.)

In \er{Pr0Qr0}, $\dot{\alpha}$ is zero at an origin where $W = 1$, but non-zero at most locations.  It does not diverge when $W = 0$ since $S'/W$ must be finite at a spatial extremum \cite{HelKra08}.  Now $\theta$ is constant along an $r$ coordinate line, $S$ does not change sign anywhere, and $W \geq 0$, while $(1 - W^2)$ only changes sign at a transition between elliptic and hyperbolic regions.  
Thus, within a purely elliptic or purely hyperbolic region, the frame rotation will be continuously in the same sense as long as $S'$ does not change sign, and would certainly integrate up to a finite rotation over a finite proper distance.  The only exception is the symmetry axis \cite{GeoHel17}, $\theta = 0, \pi$, along which the ONT is indeed constant.  The relative boost is
\begin{align}
   \dot{\chi} \; b^i & = \left[ 0 , \Rt' S + \Rt S' \cos\theta , - \Rt S' W \sin\theta , 0 \right] \frac{v^r}{S W}
\end{align}
and it has $r$ and $\theta$ components.  (It may be compared with \er{chi-dot-Pr0Qr0H} \& \er{chi-dot-Pr0Qr0P} for $\epsilon = -1,0$.)  Given the non-concentricity of the S metric, the relative boost is not too surprising.  We will not concern ourselves further with the boost rate.

For $0 = S' = Q'$ the rotation is more complex
\begin{align}\begin{aligned}
   \dot{\alpha} a^i & = \left[ 0, - \sin\theta \sin\phi , \frac{- \{1 - W^2 (1 - \cos\theta)\} \sin\phi}{W} ,
      - \frac{\{(1 - \cos\theta) W^2 + \cos\theta\} \cos\phi}{W} \right] \frac{P'}{S} v^r
   \showlabel{alphadotai-Sr0Qr0}
\end{aligned}\end{align}
(c.f. \er{alpha-dot-Sr0Qr0H} \& \er{alpha-dot-Sr0Qr0P}).  Again, $\theta$ \& $\phi$ are constant, and $S$ \& $W$ do not change sign, along an $r$ coordinate line.  So if $P'$ does not change sign, there will be a systematic rotation about the $\be_{(n)}$ axis, that is a systematic change in ${\cal V}^{(\phi)}{}_{(\theta)}$, if not the other two axes and planes as well.  In this case there is no $\theta$ value that makes all the components of \er{alphadotai-Sr0Qr0} zero.

The case $0 = S' = P'$ is of course rather similar,
\begin{align}\begin{aligned}
   \dot{\alpha} a^i & = \left[ 0, \sin\theta \cos\phi , \frac{\{1 - W^2 (1 - \cos\theta)\} \cos\phi}{W} ,
      - \frac{\{(1 - \cos\theta) W^2 + \cos\theta\} \sin\phi}{W} \right] \frac{Q'}{S} v^r
\end{aligned}\end{align}
(c.f. \er{alpha-dot-Sr0Pr0H} \& \er{alpha-dot-Sr0Pr0P}), and equivalent comments apply here.

In general, then, we only get zero rotation along $r$ coordinate paths if  $S' = 0 = P' = Q'$.

It is clear that \er{Vphth-thph} can be solved to give a path along which there is no $\theta$-$\phi$ rotation, and  \er{Vthn-thph} \& \er{Vphn-thph} can be solved to give a path along which there is no $r$-$\theta$ or $r$-$\phi$ rotation, but there is no path through a general point with $v^r \neq 0$ that preserves tetrad orientation.

\section{Conclusions}

We questioned whether the 2-spheres of constant $(t, r)$ in Szekeres models have constant orientation.  Consequently, we defined an ortho-normal tetrad (ONT), and looked at how it changed along various paths, especially the locus of constant $r$.  As noted above, a curved space or spacetime cannot have a globally constant vector field, but a systematic rotation of the frame, and the absence of a path along which there is no rotation are both significant results.

(i).~ Within any given constant time 3-space, we found that movements in the $\theta$ and $\phi$ directions give rotations that can be understood by comparison with a 3-sphere (or 3-hyperbolid, or 3-plane), but for movements in the $r$ direction there is frame rotation in general.  This rotation can be monotonically increasing in each of the special cases considered above, for example in any region where $P' = 0 = Q'$, and $f$ \& $S'$ do not change sign.

(ii).~ If $S'$, $P'$ \& $Q'$ are general, there is no path along which the full ONT does not rotate.

(iii).~ The ONT rotation is zero only in the spherically symmetric case, $S' = 0 = P' = Q'$, or along the symmetry axis in axi-symmetric models.

(iv).~ The time evolution does not cause any change in frame orientation along the particle worldlines.

The existence of some variation of an ONT in an inhomogeneous, symmetry-free, curved spacetime is hardly surprising.  However, the absence of {\it any} spatial path along which the ONT doesn't rotate is more than a little surprising.  But the fact that such variation can increase monotonically is truly unexpected.  Even in an axi-symmetric case, only the symmetry axis has an unchanging ONT.  This is in contrast to the LT model, example (d) of appendix \ref{ONTVarExmpls}, for which the ONT is unchanging along radial paths, despite the curvature of the 3-spaces.

The concept of the spatial sections of quasi-spherical Szekeres models as a sequence of ``non-concentric spheres" has become very well established, and is an excellent intuitive first description.  The idea of displaced centres has been used with the quasi-hyperboloidal S models too.  However, as example (e) of appendix \ref{ONTVarExmpls} indicates, merely displacing coordinate spheres in flat space has no effect on ONT orientation.  Thus the results presented here indicate that there is another layer of sublety in the geometry of the model, that has not been hitherto appreciated.  ``Non-concentricity" is not the whole story.

Whenever a plot of say the density variation over some 2-d spatial slice is required, one must decide on a mapping from the Szekeres coordinates to points on the paper.  The 3-d spatial sections of the Szekeres metric are conformally flat \cite{BeEaOl77} but not flat, so this inevitably requires some distortion.  However, if there is a systematic rotation, then the mapping is more complex than has been understood.

Hitherto it seems to have been tacitly assumed that the $(\theta, \phi)$ coordinates on the 2-spheres, as defined by \er{RmPrS}-\er{RmPrH}, all have the same orientation in some sense.  Thus the fact that they don't, and especially that the rotation can increase systematically with $r$, is unexpected, and adds to the difficulty of representing sections through the S spacetime on flat paper.  The question of how to handle this issue when producing graphics requires further investigation.

A question for future research is whether one can find a frame or coordinate system for the Szekeres metric that has a less variable orientation property.  For example can one define a conformally cartesian frame that has a more stable orientation?




\appendix

\section{Finding the ONT for a Given Metric}
\showlabel{FndONT}

Finding an ONT for a diagonal metric, or one with a single off-diagonal component is easy, but not so the general case.  Let the basis vectors be $\be_{(a)}$ with coordinate components $e_{(a)}{}^i$; basis indices will be between brackets.  We suppose the space has $n$ dimensions, the coordinate metric $g_{m\ell}$ is given, and the basis metric is necessarily the Lorenzian or Euclidean Cartesian metric $\eta^{(a)(b)} = \eta_{(a)(b)}$.

It is actually more convenient to work with the covariant components of the basis vectors, $e_{(a)}{}_m$.  The set of equations to be solved is
\begin{align}
   e_{(a)}{}_m \; \eta^{(a)(b)} \; e_{(b)}{}_\ell = g_{m\ell}   \showlabel{ONT}
\end{align}
with the $n(n+1)/2$ components of $g_{m\ell}$ given, and the $n^2$ components of $(e_{(a)})_m$ to be determined.  This allows rotating and/or boosting the ONT.  If the orientation of the ONT is not important, then an easy choice is to make $(e_{(a)})_m$ upper triangular or lower triangular, so that the number of unknowns and constraints is the same,
\begin{align}
   (e_{(a)})_m = 0,~~ m > a ~~~~~~\mbox{or}~~~~~~ (e_{(a)})_m = 0,~~ m < a ~.
\end{align}
For example, in 3d, with a Euclidean metric, we calculate%
\footnote{\sf This prescription ensures $\be_{(3)}$ lies along constant $x^1$ \& $x^2$, while $\be_{(2)}$ lies along constant $x^1$ --- see the following constraint equations.}
\begin{align*}
   (e_{(a)})_m & = \begin{pmatrix} A & 0 & 0 \\ D & E & 0 \\ G & H & J \end{pmatrix}
      ~~~~\mbox{and}~~~~ (e_{(a)})_m \; \delta^{(a)(b)} \; (e_{(b)})_n = g_{m\ell} \\
   \to~~~~~~~~ (e_{(a)})_m & = \begin{pmatrix}
      \sqrt{g_{11} - \dfrac{g_{12}^2 g_{33} - 2 g_{12} g_{13} g_{23} + g_{22} g_{13}^2}{g_{33} g_{22} - g_{23}^2}}\;
         & 0 & 0 \\[1mm]
      \dfrac{g_{12} g_{33} - g_{13} g_{23}}{\sqrt{g_{33} (g_{33} g_{22} - g_{23}^2)}\;}
         & \sqrt{\dfrac{g_{33} g_{22} - g_{23}^2}{g_{33}}}\; & 0 \\[6mm]
      \dfrac{g_{13}}{\sqrt{g_{33}}\;} & \dfrac{g_{23}}{\sqrt{g_{33}}\;} & \sqrt{g_{33}}\;
   \end{pmatrix} \\
\end{align*}

If in addition we wish to specify the directions of the ON basis vectors, we have further constraints.  We can specify $(n-1)$ components of the first basis vector (BV), $(n-2)$ of the second BV, and so on, down to $1$ component of the $(n-1)$th BV, and no components of the $n$th BV; i.e. $n(n-1)$ constraints.  Some possible constraints, for any given basis vector $\be_{(A)}$, are given below.

If we say that a particular $\be_{(A)}$ lies in the planes of constant $f$, then
\begin{align}
   \be_{(A)}(\bd f) = 0 = e_{(A)}{}^o \bp_o (\p_n f) \bd x^n = e_{(A)}{}^n f_{,n} = e_{(A)}{}_m g^{mn} f_{,n} ~.
\end{align}
If we specify that $\be_{(A)}$ lies along the $x^i$ coordinate lines, then it lies in the ``level planes" of all the other coordinates
\begin{align}
   j \neq i:~~ \be_{(A)}(\bd x^j) = 0 = e_{(A)}{}^j = e_{(A)}{}_m g^{mj} ~.
\end{align}
If we require that $\be_{(A)}$ is orthogonal to vector $\mb v$ then
\begin{align}
   \be_{(A)} \cdot {\mb v} = 0 = e_{(A)}{}^j g_{jk} v^k = e_{(A)}{}_k v^k ~.
\end{align}
If we want to make $\be_{(A)}$ parallel to vector $\mb v$ then
\begin{align}
   \be_{(A)} = N {\mb v} ~~~~\to~~~~ & e_{(A)}{}_i e^{(b)}{}^i = \delta^b_A = N v^i e^{(b)}{}_i = N v^i \eta^{(b)(c)} e_{(c)}{}_i \nn \\
   \therefore~~~~~~~~ & b \neq A:~~ v^i \eta^{(b)(c)} e_{(c)}{}_i = 0 ~.
\end{align}

These results are used in section \S\ref{SzAngONT} to obtain an ONT with a particular orientation.  For a general basis, use $g_{(a)(b)}$ instead of $\eta_{(a)(b)}$ in the above equations.

\section{Rotations, Boosts \& Axes}
\showlabel{RBA}

Given a 3-d rotation matrix $\mb R$ in flat space with cartesian coordinates, then a vector $\mb a$ along the axis of rotation will be unaffected by the rotation, so a unit vector giving the axis orientation is found by solving the eigenvector equation plus the unit vector condition,
\begin{align}
   \mb{R a = a ~,~~~~~~ a \cdot a} = 1 ~.
\end{align}
If instead we have a rate of rotation matrix $\mb{\dot{R}}$, then we get the axis unit vector from
\begin{align}
   \mb{\dot{R} a = 0 ~,~~~~~~ a \cdot a} = 1 ~.
\end{align}
We take such a rate of rotation matrix as always being relative to the current orientation of the given frame.  
To get the rotation rate, we find a unit vector $\mb w$ orthogonal to $\mb a$,
\begin{align}
   \mb{a \cdot w} = 0 ~,~~~~~~ \mb{w \cdot w} = 1 ~,
\end{align}
and calculate
\begin{align}
   \dot{\alpha} = \mb{a \cdot (w \times \dot{R} w)} ~.
\end{align}
In practice, this leads to 
\begin{align}
   \dot{\alpha} = \sqrt{\big(\dot{R}_{21}\big)^2 + \big(\dot{R}_{31}\big)^2 + \big(\dot{R}_{32}\big)^2}\; ~,~~~~~~
   \dot{\alpha} {\mb a} = \begin{pmatrix} R_{32} \\ -R_{31} \\ R_{21} \end{pmatrix} ~.
\end{align}
Conversely, given an axis unit vector $\mb a$, with $(a^x)^2 + (a^y)^2 + (a^z)^2 = 1$, and an angle $\alpha$, then the rotation matrix is
\begin{align}
   \mb{R} = 
   \begin{pmatrix}
   \cos\alpha + (a^x)^2 (1 - \cos\alpha)
        & a^x a^y (1 - \cos\alpha) - a^z \sin\alpha
             & a^x a^z (1 - \cos\alpha) + a^y \sin\alpha \\
   a^y a^x (1 - \cos\alpha) + a^z \sin\alpha
        & \cos\alpha + (a^y)^2 (1 - \cos\alpha)
             & a^y a^z (1 - \cos\alpha) - a^x \sin\alpha \\
   a^z a^x (1 - \cos\alpha) - a^y \sin\alpha
        & a^z a^y (1 - \cos\alpha) + a^x \sin\alpha
             & \cos\alpha + (a^z)^2 (1 - \cos\alpha)
   \end{pmatrix}
   \showlabel{GenRotMx}
\end{align}
and its rate of rotation matrix is found by differentiating and setting $\alpha$ to zero,
\begin{align}
   \mb{\dot{R}} = \dot{\alpha}
   \begin{pmatrix}
   0 & - a^z & a^y \\
   a^z & 0 & - a^x \\
   - a^y & a^x & 0
   \end{pmatrix} ~.
\end{align}
A sequence of rotations cannot in general be resolved as a rotation about a single axis in the form \er{GenRotMx}.

Given a 4-d boost matrix $B^c{}_d$ in flat spacetime with cartesian coordinates, then a spatial vector $s^d$ perpendicular to the direction (axis) of the boost will be unaffected by the boost.  More usefully, the boost matrix acting on a pure time vector will add spatial components in the 3-d direction $b^a$ of the boost,
\begin{align}
   B^c{}_d \delta^d_0 =
      \begin{pmatrix}
      \gamma \\ \gamma\beta b^x \\ \gamma\beta b^y \\ \gamma\beta b^z
      \end{pmatrix}
      ~,~~~~~~ \gamma = \cosh\chi ~,~~ \beta = \tanh\chi ~.
\end{align}
If instead you have a rate of boost matrix $\dot{B}^c{}_d$, then making it act on a time unit vector immediately gives a multiple of the spatial boost direction vector,
\begin{align}
   U^c = \dot{B}^c{}_d \delta^d_0 = \dot{\chi}
      \begin{pmatrix}
      0 \\ b^x \\ b^y \\ b^z
      \end{pmatrix} ~.
\end{align}
Similarly, the rate of boost matrix acting on that vector gives a pure time vector
\begin{align}
   \dot{B}^a{}_c U^c = \dot{\chi}
      \begin{pmatrix}
      (b^x)^2 + (b^y)^2 + (b^z)^2 \\ 0 \\ 0 \\ 0
      \end{pmatrix}
      ~,~~~~~~ (b^x)^2 + (b^y)^2 + (b^z)^2 = 1 ~,
\end{align}
so that the rate of boost is clearly the dot product with a unit time vector
\begin{align}
   \dot{\chi} = \delta_a^0 U^a = \delta_a^0 \dot{B}^a{}_c \dot{B}^c{}_d \delta^d_0 = \dot{B}^0{}_c \dot{B}^c{}_0
   ~,~~~~~~ \dot{\beta} = \dot{\chi} \, \big[ {\rm sech}^2\chi \big]_{\chi = 0} = \dot{\chi} ~,
\end{align}
since the rate of boost is relative to the current frame.  In terms of the given rate of boost matrix, this evaluates to
\begin{align}
   \dot{\beta} = \dot{\chi} = \sqrt{\big(\dot{B}_{01}\big)^2 + \big(\dot{B}_{02}\big)^2 + \big(\dot{B}_{03}\big)^2}\; ~,~~~~~~
   \dot{\chi} {\mb b} = \begin{pmatrix} 0 \\ B_{12} \\ B_{13} \\ B_{14} \end{pmatrix} ~.
\end{align}
Conversely, given a unit 3-vector $\mb b$ along the boost axis, and a boost speed $\beta = \tanh\chi$, then the boost matrix is
\begin{align}
   B^c{}_d =
   \begin{pmatrix}
   \cosh\chi & \sinh\chi \, b^x & \sinh\chi \, b^y & \sinh\chi \, b^z \\
   \sinh\chi \, b^x & 1 + (\cosh\chi - 1) \, (b^x)^2 & (\cosh\chi - 1) \, b^x b^y & (\cosh\chi - 1) \, b^x b^z \\
   \sinh\chi \, b^y & (\cosh\chi - 1) \, b^y b^x & 1 + (\cosh\chi - 1) \, (b^y)^2 & (\cosh\chi - 1) \, b^y b^z \\
   \sinh\chi \, b^z & (\cosh\chi - 1) \, b^z b^x & (\cosh\chi - 1) \, b^z b^y & 1 + (\cosh\chi - 1) \, (b^z)^2
   \end{pmatrix} ~,
   \showlabel{GenBstMx}
\end{align}
and its rate of boost matrix is
\begin{align}
   \dot{B}^c{}_d = \dot{\chi}
   \begin{pmatrix}
   0 & b^x & b^y & b^z \\
   b^x & 0 & 0 & 0 \\
   b^y & 0 & 0 & 0 \\
   b^z & 0 & 0 & 0
   \end{pmatrix} ~.
\end{align}
A sequence of boosts cannot in general be resolved as a boost along a single axis in the form \er{GenBstMx}.

A general Lorentz transformation $\Lambda^c{}_d$, is a combination of rotations and boosts, that is not easily decomposed; it should have determinant 1.  With the general rate of Lorentz transformation $\dot{\Lambda}^c{}_d$, however, the symmetric part (in the matrix sense) is the rate of boost, and the anti-symmetric part is the rate of rotation,
\begin{align}
   {}^s\!\dot{\Lambda}^c{}_d & = \frac{1}{2} \left( \dot{\Lambda}^c{}_d + (\dot{\Lambda}^T)^c{}_d \right)
      = \mbox{rate of boost matrix} ~, \\
   {}^a\!\dot{\Lambda}^c{}_d & = \frac{1}{2} \left( \dot{\Lambda}^c{}_d - (\dot{\Lambda}^T)^c{}_d \right)
      = \mbox{rate of rotation matrix} ~,
\end{align}
each of which can be treated as above.  Spatial 3-vectors just need a zero for the time component.
Note that $\dot{\Lambda}^c_d$ does not have zero determinant in general, but each of ${}^s\!\dot{\Lambda}^c_d$ \& ${}^a\!\dot{\Lambda}^c_d$ does.

\subsection{Examples}
\showlabel{ONTVarExmpls}

The interpretation of results in the main text will be aided by considering a class of examples.

(a)~~First consider 3-d flat space with spherical coordinates.  The ONT aligned with the spherical coordinates, and the consequent path-rotation-rate matrix, are
\begin{align}
   e_{(a)}{}_i = \begin{pmatrix}
   1 & 0 & 0 \\
   0 & r & 0 \\
   0 & 0 & r \sin\theta
   \end{pmatrix} ~,~~~~~~~~
   {\cal V}^{(c)}{}_{(b)} = \begin{pmatrix}
   0 & - v^\theta & - \sin\theta \; v^\phi \\
   v^\theta & 0 & - \cos\theta \; v^\phi \\
   \sin\theta \; v^\phi & \cos\theta \; v^\phi & 0
   \end{pmatrix} ~.
\end{align}
From these, we find the rotation rate and axis are
\begin{align}
   \dot{\alpha} & = \sqrt{(v^\theta)^2 + (v^\phi)^2}\; ~,~~~~~~~~
   \dot{\alpha} \; a^i = \begin{pmatrix}
   \cos\theta \; v^\phi \\
   - \sin\theta \; v^\phi \\
   v^\theta
   \end{pmatrix} ~.
\end{align}
For motion in the $\theta$ direction, $v^r = 0 = v^\phi$, there is a rotation rate of $v^\theta$ about the local $\be_{(\phi)}$ axis.  For motion in the $\phi$ direction, $v^r = 0 = v^\theta$, there is a rotation rate of $v^\phi$ about the ``$z$" axis, i.e. a direction parallel to $\theta = 0$.
Since there is no dependence on $v^r$, it is evident that the ONT has constant orientation along $r$ coordinate lines.

(b)~~Next consider the 3-sphere,
\begin{align}
   e_{(a)}{}_i = \begin{pmatrix}
   1 & 0 & 0 \\
   0 & \sin\psi & 0 \\
   0 & 0 & \sin\psi \sin\theta
   \end{pmatrix} ~,~~~~~~~~
   {\cal V}^{(c)}{}_{(b)} = \begin{pmatrix}
   0 & - \cos\psi \; v^\theta & - \cos\psi \sin\theta \; v^\phi \\
   \cos\psi \; v^\theta & 0 & - \cos\theta \; v^\phi \\
   \cos\psi \sin\theta \; v^\phi & \cos\theta \; v^\phi & 0
   \end{pmatrix} ~,
\end{align}
from which the rotation rate and axis are,
\begin{align}
   \dot{\alpha} & = \sqrt{(v^\theta)^2 \; \cos^2\psi + (v^\phi)^2 (1 - \sin^2\psi \sin^2\theta)}\; ~,~~~~~~~~
   \dot{\alpha} \; a^i = \begin{pmatrix}
   \cos\theta \; v^\phi \\
   - \cos\psi \sin\theta \; v^\phi \\
   \cos\psi \; v^\theta
   \end{pmatrix} ~.
\end{align}
For $\psi \approx 0$, this looks like the above result, and in particular there is no ONT rotation along constant $\psi$ lines.  However, the surface $\psi = \pi/2$ is the constant-$\psi$ 2-sphere of maximum areal radius.  On this surface, the $\theta$ lines are geodesic, so $v^\theta$ generates no rotation.  Similarly the $\phi$ line at $\theta = \pi/2$ is geodesic, but the other $\phi$ lines aren't.  At intermediate $\psi$ values, $v^\theta$ still causes rotation about the local $\be_{(\phi)}$ axis by a reduced amount, and $v^\phi$ causes reduced rotation about an axis that varies from the ``$z$" direction at $\psi = 0$ to the ``radial" direction at $\psi = \pi/2$.  The areal radius of each constant-$\psi$ 2-sphere is $\sin\psi$; this is what's used in an embedding.  However, the effective radius is $1/\cos\psi$, which represents how much the ONT is rotating relative to a geodesic.%
\footnote{\sf This illustrates one of the subtleties of curved spaces that is inherited by GR; although all the constant $\psi$ 2-spheres are intrinsically curved, as part of a 3-space, one or more of them may have zero extrinsic curvature.  Whether or not a direction is considered constant depends on which space it lives in.}

(c)~~We re-do the 3-sphere, except we rotate the $\phi$ coordinate as $\psi$ increases: $\phi = \tilde{\phi} + \omega \tilde{\psi}$.  Discarding the tilde, the basis and rotation rate matrices are
\begin{align}\begin{aligned}
   e_{(a)}{}_i & = \begin{pmatrix}
   1 & 0 & 0 \\
   0 & \sin\psi & 0 \\
   \omega \sin\psi \sin\theta & 0 & \sin\psi \sin\theta
   \end{pmatrix} ~, \\
   {\cal V}^{(c)}{}_{(b)} & = \begin{pmatrix}
   0 & - \cos\psi v^\theta & - \cos\psi \sin\theta (v^\phi + \omega \;v^r) \\
   \cos\psi v^\theta & 0 & - \cos\theta (v^\phi + \omega \;v^r) \\
   \cos\psi \sin\theta (v^\phi + \omega \;v^r) & \cos\theta (v^\phi + \omega \;v^r) & 0
   \end{pmatrix} ~.
\end{aligned}\end{align}
Now the rotation rate and axis are
\begin{align}
   \dot{\alpha} & = \sqrt{(v^\theta)^2 + (v^\phi)^2}\; ~,~~~~~~~~
   \dot{\alpha} \; a^i = \begin{pmatrix}
   \cos\theta (v^\phi + \omega \;v^r) \\
   - \cos\psi \sin\theta (v^\phi + \omega \;v^r) \\
   \cos\psi \; v^\theta
   \end{pmatrix} ~.
\end{align}

(d)~~Consider the {\LT} model, with $R = R(t,r)$, $W = W(r)$, ${}' = \pdil{}{r}$, $\dot{} = \pdil{}{t}$,
\begin{align}\begin{aligned}
   e_{(a)}{}_i = \begin{pmatrix}
   -1 & 0 & 0 & 0 \\
   0 & R'/W & 0 & 0 \\
   0 & 0 & R & 0 \\
   0 & 0 & 0 & R \sin\theta
   \end{pmatrix} ~, \\
   {\cal V}^{(c)}{}_{(b)} = \begin{pmatrix}
   0 & (\Rt'/W) \; v^r & \Rt \; v^\theta & \Rt \sin\theta \; v^\phi \\
   (\Rt'/W) \; v^r & 0 & - W \; v^\theta & - W \sin\theta \; v^\phi \\
   \Rt \; v^\theta & W \; v^\theta & 0 & - \cos\theta \; v^\phi \\
   \Rt \sin\theta \; v^\phi & W \sin\theta \; v^\phi & \cos\theta \; v^\phi & 0
   \end{pmatrix} ~.
   \showlabel{eVLT}
\end{aligned}\end{align}
In this case, the rotation rate and axis, are,
\begin{align}
   \dot{\alpha} & = \sqrt{(v^\theta)^2 \; W^2 + (v^\phi)^2 (1 - (1 - W^2) \sin^2\theta)}\; ~,~~~~~~~~
   \dot{\alpha} \; a^i = \begin{pmatrix}
   0 \\
   \cos\theta \; v^\phi \\
   - W \sin\theta \; v^\phi \\
   W \; v^\theta
   \end{pmatrix} ~.
   \showlabel{rotrateaxLT}
\end{align}
We get essentially the same interpretation as for the 3-sphere, if we let $R$ play the role of $\sin\psi$, and $W$ the role of $\cos\psi$, except that the {\LT} model is radially inhomogeneous.  In particular, the ONT is constant in the radial direction.  The sphere of maximum or minimum areal radius, if there is one, is at $r$ values where $W = 0 = R'$ for all time.  The addition of a time dimension means we also have a boost rate and boost axis,
\begin{align}
   \dot{\chi} & = \sqrt{(v^\theta)^2 \; \cos^2\psi + (v^\phi)^2 (1 - \sin^2\psi \sin^2\theta)}\; ~,~~~~~~~~
   \dot{\chi} \; b^i = \begin{pmatrix}
   0 \\
   (\Rt'/W) \; v^r \\
   \Rt \; v^\theta \\
   \Rt \sin\theta \; v^\phi
   \end{pmatrix} ~.
   \showlabel{bstrateaxLT}
\end{align}
Here the interpretation is easy.  Thinking of the expansion as radial, then movements in the $v^r$, $v^\theta$ \& $v^\phi$ directions give the expected relative boost between two radial velocities.

(e)~~Non-concentric spherical coordinates in 3-d flat space.  This is the displaced-spheres version of example (a), with the centre of the sphere at radius $r$ located at $Z(r)$ on the $z$ axis, $\theta = 0$.  The ONT is
\begin{align}\begin{aligned}
   e_{(a)}{}_i = \begin{pmatrix}
   1 + Z' \cos\theta & 0 & 0 \\
   - Z' \sin\theta & r & 0 \\
   0 & 0 & r \sin\theta
   \end{pmatrix} ~,
   \showlabel{eDS}
\end{aligned}\end{align}
but the ${\cal V}^{(c)}{}_{(b)}$ is identical to that of example (a).  In flat space, the basis orientation is unaffected by the displacement of the coordinate spheres.

\section{Key Equations for $\epsilon = -1$}
\showlabel{AppEp-1}

By \er{RmPrH}, the $\epsilon = -1$ metric in $(\theta, \phi)$ coordinates is
\begin{align}
   g_{tt} & = - 1 ~,~~~~
   g_{\theta\theta} = R^2 ~,~~~~
   g_{\phi\phi} = R^2 \sinh^2\theta ~, \nn \\
   g_{rr} & = \frac{\big(R' S + R [S' \cosh\theta + \sinh\theta \{P' \cos\phi + Q' \sin\phi\}]\big)^2}{S^2 W^2} \nn \\
      &~~~~ + \frac{R^2}{S^2} \big(S' \sinh\theta + (\cosh\theta - 1) \{P' \cos\phi + Q' \sin\phi\}\big)^2 \nn \\
      &~~~~ + \frac{R^2}{S^2} (\cosh\theta - 1)^2 \{P' \sin\phi - Q' \cos\phi\}^2 ~, \nn \\
   g_{r\theta} & = - \frac{R^2 (S' \sinh\theta + (\cosh\theta - 1) \{P' \cos\phi + Q' \sin\phi\})}{S} \nn \\
   g_{r\phi} & = - \frac{R^2 (\cosh\theta - 1) \sinh\theta \{P' \sin\phi - Q' \cos\phi\}}{S} ~,
   \showlabel{SzThPhMetricH}
\end{align}
and the functions $E$ and $E'$ become
\begin{align}
   E = \frac{S}{(\cosh\theta - 1)} ~,~~~~~~ 
   E' = - \frac{\cosh\theta S' + \sinh\theta (P' \cph + Q' \sph)}{(\cosh\theta - 1)} ~.
\end{align}
The non-zero components of the corresponding ONT are
\begin{align}
   e_{(a)}{}_i & = 
   \begin{pmatrix}
   -1 & 0 & 0 & 0 \\
   0 & \dfrac{1}{W} \left( R' + \dfrac{R [S' \cosh\theta + \sinh\theta \{P' \cos\phi + Q' \sin\phi\}]}{S} \right) & 0 & 0 \\[4mm]
   0 & - \dfrac{R [S' \sinh\theta + (\cosh\theta - 1) \{P' \cos\phi + Q' \sin\phi\}]}{S} & R & 0 \\[3mm]
   0 & - \dfrac{R (\cosh\theta - 1) \{P' \sin\phi - Q' \cos\phi\}}{S} & 0 & R \sinh\theta
   \end{pmatrix}
   \showlabel{SzThPhONTH}
\end{align}
The variation of this ONT for $\epsilon = -1$ in terms of the $v^j$ is given by
\begin{align}
   {\cal V}^{(n)}{}_{(t)} & = \frac{(\Rt' S + \Rt [S' \cosh\theta + \sinh\theta \{P' \cos\phi + Q' \sin\phi\}])}
       {S W} \; v^r   \showlabel{Vnt-thphH} \\
   {\cal V}^{(\theta)}{}_{(t)} & = \frac{- \Rt [S' \sinh\theta + (\cosh\theta - 1)\{P' \cos\phi + Q' \sin\phi\}]}{S} \; v^r
      + \Rt \; v^\theta   \showlabel{Vtht-thphH} \\
   {\cal V}^{(\phi)}{}_{(t)} & = \frac{- \Rt (\cosh\theta - 1)\{P' \sin\phi - Q' \cos\phi\}}{S} \; v^r
      + \Rt \sinh\theta \; v^\phi   \showlabel{Vpht-thphH} \\
   {\cal V}^{(\theta)}{}_{(n)} & = \bigg( \frac{S' (W^2 + 1) \sinh\theta + [\cosh\theta (W^2 + 1) - W^2]
      \{P' \cos\phi + Q' \sin\phi\}}{W S} \bigg) v^r - W v^\theta   \showlabel{Vthn-thphH} \\
   {\cal V}^{(\phi)}{}_{(n)} & = \frac{[W^2 (\cosh\theta - 1) - 1]\{P' \sin\phi - Q' \cos\phi\}}{S W} \; v^r
      - W \sinh\theta \; v^\phi   \showlabel{Vphn-thphH} \\
   {\cal V}^{(\phi)}{}_{(\theta)} & = \frac{\sinh\theta \{P' \sin\phi - Q' \cos\phi\}}{S} \; v^r
      - \cosh\theta v^\phi   \showlabel{Vphth-thphH}
\end{align}

Along a pure $r$ direction ($0 = v^t = v^\theta = v^\phi$, and $v^r = 1$) we have the following special cases.  
When $0 = P' = Q'$, the relative rotation is
\begin{align}\begin{aligned}
   \dot{\alpha} & = - \frac{S' (1 + W^2) \sinh\theta}{S W} v^r ~,~~~~~~
   a^i = \delta^i_\phi ~,   \showlabel{alpha-dot-Pr0Qr0H}
\end{aligned}\end{align}
and the relative boost is
\begin{align}
   \dot{\chi} \; b^i & = \left[ 0 , \Rt' S + \Rt S' \cosh\theta , - \Rt S' W \sinh\theta , 0 \right] \frac{v^r}{S W} ~.
   \showlabel{chi-dot-Pr0Qr0H}
\end{align}
When $0 = S' = Q'$, the relative rotation is
\begin{align}\begin{aligned}
   \dot{\alpha} a^i & = \left[ 0, - \sinh\theta \sin\phi , \frac{- \{1 - W^2 (\cosh\theta - 1)\} \sin\phi}{W} ,
      - \frac{\{(\cosh\theta - 1) W^2 + \cosh\theta\} \cos\phi}{W} \right] \frac{P'}{S} v^r ~,
   \showlabel{alpha-dot-Sr0Qr0H}
\end{aligned}\end{align}
and when $0 = S' = P'$ it is
\begin{align}\begin{aligned}
   \dot{\alpha} a^i & = \left[ 0, \sinh\theta \cos\phi , \frac{\{1 - W^2 (\cosh\theta - 1)\} \cos\phi}{W} ,
      - \frac{\{(\cosh\theta - 1) W^2 + \cosh\theta\} \sin\phi}{W} \right] \frac{Q'}{S} v^r ~.
   \showlabel{alpha-dot-Sr0Pr0H}
\end{aligned}\end{align}

\section{Key Equations for $\epsilon = 0$}
\showlabel{AppEp0}

The $\epsilon = 0$ metric in $(\theta, \phi)$ coordinates is, by \er{RmPrP}, 
\begin{align}
   g_{tt} & = - 1 ~,~~~~
   g_{\theta\theta} = R^2 ~,~~~~
   g_{\phi\phi} = R^2 \theta^2 ~, \nn \\
   g_{rr} & = \frac{\big(R' S + R [S' + \theta \{P' \cos\phi + Q' \sin\phi\}]\big)^2}{S^2 W^2} \nn \\
      &~~~~ + \frac{R^2}{S^2} \big(S' \theta + (\theta^2/2) \{P' \cos\phi + Q' \sin\phi\}\big)^2 \nn \\
      &~~~~ + \frac{R^2}{S^2} (\theta^2/2)^2 \{P' \sin\phi - Q' \cos\phi\}^2 ~, \nn \\
   g_{r\theta} & = - \frac{R^2 (S' \theta + (\theta^2/2) \{P' \cos\phi + Q' \sin\phi\})}{S} \nn \\
   g_{r\phi} & = - \frac{R^2 (\theta^3/2) \{P' \sin\phi - Q' \cos\phi\}}{S} ~,
   \showlabel{SzThPhMetricP}
\end{align}
while the functions $E$ and $E'$ become
\begin{align}
   E = \frac{2 S}{\theta^2} ~,~~~~~~ 
   E' = - \frac{2 [S' + \theta (P' \cph + Q' \sph)]}{\theta^2} ~.
\end{align}
For this $\epsilon$ value, the ONT (non-zero components) is
\begin{align}
   e_{(a)}{}_i & = 
   \begin{pmatrix}
   -1 & 0 & 0 & 0 \\
   0 & \dfrac{1}{W} \left( R' + \dfrac{R [S' + \theta \{P' \cos\phi + Q' \sin\phi\}]}{S} \right) & 0 & 0 \\[4mm]
   0 & - \dfrac{R [S' \theta + (\theta^2/2) \{P' \cos\phi + Q' \sin\phi\}]}{S} & R & 0 \\[3mm]
   0 & - \dfrac{R (\theta^2/2) \{P' \sin\phi - Q' \cos\phi\}}{S} & 0 & R \theta
   \end{pmatrix}
   \showlabel{SzThPhONTP}
\end{align}

The ONT variation in terms of the $v^j$, when $\epsilon = 0$, is
\begin{align}
   {\cal V}^{(n)}{}_{(t)} & = \frac{(\Rt' S + \Rt [S' + \theta \{P' \cos\phi + Q' \sin\phi\}])}
       {S W} \; v^r   \showlabel{Vnt-thphP} \\
   {\cal V}^{(\theta)}{}_{(t)} & = \frac{- \Rt [S' \theta + (\theta^2/2)\{P' \cos\phi + Q' \sin\phi\}]}{S} \; v^r
      + \Rt \; v^\theta   \showlabel{Vtht-thphP} \\
   {\cal V}^{(\phi)}{}_{(t)} & = \frac{- \Rt (\theta^2/2)\{P' \sin\phi - Q' \cos\phi\}}{S} \; v^r
      + \Rt \theta \; v^\phi   \showlabel{Vpht-thphP} \\
   {\cal V}^{(\theta)}{}_{(n)} & = \bigg( \frac{S' W^2 \theta + [W^2 (\theta^2/2) + 1]
      \{P' \cos\phi + Q' \sin\phi\}}{W S} \bigg) v^r - W v^\theta   \showlabel{Vthn-thphP} \\
   {\cal V}^{(\phi)}{}_{(n)} & = \frac{[W^2 (\theta^2/2) - 1]\{P' \sin\phi - Q' \cos\phi\}}{S W} \; v^r
      - W \theta \; v^\phi   \showlabel{Vphn-thphP} \\
   {\cal V}^{(\phi)}{}_{(\theta)} & = \frac{\theta \{P' \sin\phi - Q' \cos\phi\}}{S} \; v^r
      - v^\phi   \showlabel{Vphth-thphP}
\end{align}

Again considering a pure $r$ direction, $0 = v^t = v^\theta = v^\phi$, $v^r = 1$, the same set of special cases give the following.  
Putting $0 = P' = Q'$, the relative rotation is
\begin{align}\begin{aligned}
   \dot{\alpha} & = - \frac{S' W \theta}{S} v^r ~,~~~~~~
   a^i = \delta^i_\phi ~.   \showlabel{alpha-dot-Pr0Qr0P}
\end{aligned}\end{align}
and the relative boost is
\begin{align}
   \dot{\chi} \; b^i & = \left[ 0 , \Rt' S + \Rt S' , - \Rt S' W \theta , 0 \right] \frac{v^r}{S W}
   \showlabel{chi-dot-Pr0Qr0P}
\end{align}
The rotation in the $0 = S' = Q'$ case is,
\begin{align}\begin{aligned}
   \dot{\alpha} a^i & = \left[ 0, - \theta \sin\phi , \frac{- \{1 - W^2 (\theta^2/2)\} \sin\phi}{W} ,
      - \frac{\{1 + W^2 (\theta^2/2)\} \cos\phi}{W} \right] \frac{P'}{S} v^r ~,
   \showlabel{alpha-dot-Sr0Qr0P}
\end{aligned}\end{align}
and in the $0 = S' = P'$ case it is,
\begin{align}\begin{aligned}
   \dot{\alpha} a^i & = \left[ 0, \theta \cos\phi , \frac{\{1 - W^2 (\theta^2/2)\} \cos\phi}{W} ,
      - \frac{\{1 + W^2 (\theta^2/2)\} \sin\phi}{W} \right] \frac{Q'}{S} v^r ~.
   \showlabel{alpha-dot-Sr0Pr0P}
\end{aligned}\end{align}

\end{document}